\documentclass[twocolumn]{aastex63}
\pdfoutput=1
\usepackage{amsmath,amstext}
\usepackage[T1]{fontenc}
\usepackage[figure,figure*]{hypcap}

\shorttitle{Conductive Opacities in WD Envelopes}
\shortauthors{Blouin et al.}

\begin{document}

\title{New Conductive Opacities for White Dwarf Envelopes}

\author{Simon Blouin}
\correspondingauthor{Simon Blouin}
\affiliation{Los Alamos National Laboratory, PO Box 1663, Los Alamos, NM 87545, USA}
\email{sblouin@lanl.gov}

\author{Nathaniel R. Shaffer}
\affiliation{Los Alamos National Laboratory, PO Box 1663, Los Alamos, NM 87545, USA}

\author{Didier Saumon}
\affiliation{Los Alamos National Laboratory, PO Box 1663, Los Alamos, NM 87545, USA}

\author{Charles E. Starrett}
\affiliation{Los Alamos National Laboratory, PO Box 1663, Los Alamos, NM 87545, USA}

\begin{abstract}
Thanks to their continuous cooling and relative simplicity, white dwarf stars are routinely used to measure the ages of stellar populations. The usefulness of white dwarfs as cosmochronometers depends on the availability of accurate cooling models. A key ingredient of those models are the conductive opacities, which largely govern the cooling rate. In this work, we present improved conductive opacities for the regime of moderate coupling and moderate degeneracy that characterizes an important portion of the envelopes of DA and DB white dwarfs. We find differences of up to a factor 3 between our calculations and the commonly used opacities of \cite{cassisi2007}, which we attribute to an improved account of electron--electron scattering. The cooling models are strongly affected by those changes in the conductive opacities: the age of a 4000\,K white dwarf can be reduced by as much as 2\,Gyr. We provide analytical fits to our new opacities to facilitate the implementation of this important effect in white dwarf evolution codes.
\end{abstract}

\section{Introduction}
One of the most basic properties of a star is its age, yet measuring individual stellar ages remains a challenging problem \citep{soderblom2010}. Even the Sun does not directly reveal its age: our best constraints on its age come from laboratory studies of solar system material. This difficulty limits what we can learn about time-dependent processes such as stellar evolution and the formation history of our Galaxy. 

Fortunately, stellar remnants can provide a solution to this problem. The evolution of white dwarfs is simpler than that of main-sequence stars and their monotonic cooling implies that there is a relatively simple mapping between their age and their temperature \citep{mestel1952}. Therefore, precise white dwarf ages can be obtained with the help of theoretical evolution sequences to model their cooling \citep[e.g.,][]{hansen1999,fontaine2001,salaris2010,renedo2010} and atmosphere models to measure their atmospheric parameters from comparisons with spectroscopic or photometric observations \citep[e.g.,][]{bergeron1995,dufour2005,koester2010,tremblay2013,blouin2018}. This technique allows one to precisely measure the ages of different components of the Milky Way \citep{winget1987,oswalt1996,garciaberro2010,jeffery2011,kalirai2012,hansen2013,kilic2017,kilic2019} and probe its formation history \citep{tremblay2014,fantin2019}.

While the evolution of white dwarfs is relatively simple to model, new observational data from \textit{Gaia} DR2 \citep{gaiadr2a,gaiadr2b} have recently highlighted the limitations of presently available theoretical evolution sequences. Current models fail to quantitatively reproduce the signature of core crystallization identified by \cite{tremblay2019}, miss an important mechanism that delays the cooling of massive objects \citep{cheng2019} and cannot reproduce the observed DA mass distribution \citep{kilic2020}. This disagreement with empirical data casts doubt on the evolution models' accuracy when used to measure the ages of stellar populations and motivates further work on their constitutive physics.

Conductive opacities are an ingredient of the utmost importance in white dwarf evolution models, since they largely control the rate at which cooling takes place. Until recently, the conductive opacities used in white dwarf evolution models were those of \cite{itoh1983}, \cite{mitake1984} and \cite{hubbard1969}. More modern calculations were published by \citet[see also \citealt{potekhin2015}]{cassisi2007} and significant differences with the previously available conductivities were identified, especially in the He envelope and the core. Those differences can affect the cooling ages of white dwarfs by as much as 10\% \citep[][B\'edard et al. in preparation]{salaris2013}. While suitable across a large range of density and temperature conditions, the conductive opacities of \cite{cassisi2007} are more approximate in the moderately degenerate and moderately coupled regime that characterizes a large portion of the H and He envelopes of DA and DB white dwarfs, since the $e-e$ collision rate is important but not well known at such conditions. 

In this work, we revise the conductive opacities of H and He under conditions relevant to white dwarf envelopes. Our new conductive opacities are given in Section~\ref{sec:opac}, where we also present the theoretical framework on which our new calculations are based. In Section~\ref{sec:implementation}, we discuss the implementation of those new opacities in white dwarf evolution codes and give analytic equations that can be implemented in any existing code to correct the \cite{cassisi2007} opacities in the relevant temperature and density regime. Section~\ref{sec:implications} explores the implications of our new conductive opacities for the cooling of white dwarfs. Finally, our conclusions are stated in Section~\ref{sec:conclusions}.

\section{New conductive opacities}
\label{sec:opac}

Our conductive opacities are evaluated from the mean-force quantum Landau--Fokker--Planck (qLFP) plasma kinetic theory, which is a recently developed model that is uniquely suited to the moderately coupled and moderately degenerate conditions typical of the conductive envelopes of DA and DB white dwarfs.
The model is described fully in \cite{shaffer2020b}.
The conductive opacity $\kappa_c$ is computed from the thermal conductivity $\lambda$ according to
\begin{equation}
  \kappa_c = \frac{16\sigma T^3}{3\rho \lambda},
  \label{eq:cond-to-opac}  
\end{equation}
where $\sigma$ is the Stefan--Boltzmann constant, $T$ the temperature and $\rho$ the density.
The thermal conductivity is obtained by a Chapman--Enskog solution of the qLFP kinetic equation, which is a standard Fokker--Planck-type plasma kinetic equation extended to account for Fermi--Dirac statistics of the electrons \citep{danielewicz1980,daligault2018}.

A hallmark of traditional Fokker--Planck theories is Coulomb logarithms, which crudely model how many-body phenomena such as screening affect the binary collision physics.
Standard analytic formulas for the Coulomb logarithms are valid only for weakly coupled plasmas and break down at the moderately coupled conditions relevant to the conductive envelopes of white dwarfs.
The Coulomb logarithms used in our calculations are instead based on mean-force scattering, where the collision cross-sections are computed numerically from the scattering phase shifts for the $e-i$ and $e-e$ potentials of mean force, $V^{\mathrm{mf}}_{ei}(r)$ and $V^{\mathrm{mf}}_{ee}(r)$.
The cross-sections are reduced to Coulomb logarithms according to
\begin{subequations}
  \begin{equation}
    \label{eq:lnL-ei}
    \ln\Lambda_{ei} = \frac{2 Q_2(\mu_e/k_BT) }{Q_{\frac12}(\mu_e/k_BT) Zm_e^{\frac12} \sigma_{\mathrm{RT}}} \left( \frac{ 2 k_B T}{\pi} \right)^{\frac32}
    ,
  \end{equation}
  \begin{equation}
    \label{eq:lnL-ee}
    \ln\Lambda_{ee} = \frac12 \left< \frac{\sigma^{(2)}_{ee}(k)}{\bar\sigma_{ee}(k)} \right>
    + \frac54 \operatorname{erf}\left[\left(\frac{2T}{3T_F}\right)^3\right]
    ,
  \end{equation}
\end{subequations}
where 
\begin{equation}
  \label{eq:sigma-rt}
  \sigma_{\mathrm{RT}} = -\frac{e^2\hbar^2}{3m_e^2} \int \tau_{ei}(k) k^2 f'(E) \frac{d\vec k}{4 \pi^3}
\end{equation}
is the electrical conductivity in mean-force relaxation-time approximation \citep{starrett2017},
$\tau_{ei}(k) = [n_i (\hbar k/m_e) \sigma^{(1)}_{ei}(k)]^{-1}$ is the $e-i$ relaxation time, $n_i$ is the ion number density, $f'(E)$ is the energy derivative of the Fermi--Dirac distribution (where $E = \hbar^2k^2/2m_e$),
\begin{equation}
  \label{eq:sigma-r}
  \sigma_{ij}^{(r)}(k) = \int (1 - \cos^r\theta) \, d\sigma_{ij}(\theta, k)
  ,
\end{equation}
are angular moments of the differential cross-section for $i-j$ scattering,
$\bar\sigma_{ee}(k) = \pi (e^2 m_e/2\hbar^2 k^2)^2$ is a reference cross-section,

\begin{equation}
  \label{eq:fd-int}
  Q_\nu(z) = \frac{1}{\Gamma(\nu+1)} \int_0^\infty \frac{t^\nu}{1+e^{t-z}} \,dt
\end{equation}
are Fermi--Dirac integrals, $\mu_e$ is the electron chemical potential, $T_F = \hbar^2(3\pi^2n_e)^{\frac23}/{2m_ek_B}$ is the Fermi temperature, and angle brackets denote an average with respect to the distribution of relative momenta $k$ between two electrons with Fermi--Dirac energy distributions.
See \cite{shaffer2020b} for additional details and a complete derivation.
  
The mean-force potentials model how many-body screening and correlations affect binary encounters between particles, even in strongly coupled plasmas.
In this work, they are obtained from the average-atom two-component plasma model \citep{starrett2013,starrett2017,shaffer2020a}, which is a finite-temperature density functional theory of ionic and electronic correlations in dense plasmas.
In the weakly coupled limit, the potentials of mean force reduce to Debye-screened Coulomb potentials, and so the effective Coulomb logarithms and opacities of qLFP are in good agreement with the analytic expressions in \cite{hubbard1969}.

Not all limitations of the Fokker--Planck glancing-collision approximation can be overcome using mean-force potentials.
In particular, the qLFP kinetic equation is not accurate for highly degenerate electrons ($T \lesssim 0.1 T_F$) due to a subtle interplay between the glancing-collision approximation and the Pauli exclusion principle, which causes qualitatively incorrect temperature dependence of the qLFP conductivity \citep{shaffer2020b}.
We expect that at degenerate conditions, the model of \cite{cassisi2007} is likely superior to qLFP.
For this reason, we use qLFP results only when $T > 0.1 T_F$.
Tables~\ref{tab:H} and \ref{tab:He} list the qLFP conductive opacities in this temperature range, which were calculated assuming a fully ionized plasma in all cases.
These values are the basis of the practical formulas described in Section~\ref{sec:implementation}.

\begin{deluxetable*}{crrrrrrrrrrr}
\tablecaption{Decimal logarithm of qLFP conductive opacity in $\mathrm{cm^2\,g^{-1}}$ for H. \label{tab:H}}
\tablehead{
 & \multicolumn{11}{c}{$\log\rho (\mathrm{g\,cm^{-3}})$} \\
$\log T (\mathrm{K})$ &  \multicolumn{1}{c}{$-$1.0} &   \multicolumn{1}{c}{0.0} &   \multicolumn{1}{c}{0.5} &   \multicolumn{1}{c}{1.0} &   \multicolumn{1}{c}{1.5} &   \multicolumn{1}{c}{2.0} &   \multicolumn{1}{c}{2.5} &   \multicolumn{1}{c}{3.0} &   \multicolumn{1}{c}{3.5} &    \multicolumn{1}{c}{4.0} &    \multicolumn{1}{c}{4.5} 
}
\startdata
5.00 & 4.881 & 2.839 & 1.889 &       &       &       &       &       &       &        &        \\
5.22 & 5.059 & 3.246 & 2.324 & 1.347 &       &       &       &       &       &        &        \\
5.44 & 5.225 & 3.696 & 2.737 & 1.799 &       &       &       &       &       &        &        \\
5.67 & 5.396 & 4.157 & 3.159 & 2.227 & 1.274 &       &       &       &       &        &        \\
5.89 & 5.571 & 4.383 & 3.631 & 2.642 & 1.719 & 0.754 &       &       &       &        &        \\
6.11 & 5.745 & 4.575 & 3.974 & 3.086 & 2.139 & 1.210 &       &       &       &        &        \\
6.33 & 5.916 & 4.763 & 4.171 & 3.533 & 2.561 & 1.642 & 0.700 &       &       &        &        \\
6.56 & 6.083 & 4.945 & 4.363 & 3.765 & 3.017 & 2.057 & 1.146 & 0.189 &       &        &        \\
6.78 & 6.242 & 5.121 & 4.547 & 3.961 & 3.355 & 2.488 & 1.565 & 0.646 &       &        &        \\
7.00 & 6.400 & 5.288 & 4.723 & 4.147 & 3.556 & 2.915 & 1.977 & 1.074 & 0.133 &        &        \\
7.22 & 6.674 & 5.449 & 4.893 & 4.327 & 3.748 & 3.150 & 2.414 & 1.481 & 0.575 & $-$0.377 &        \\
7.44 &       &       &       & 4.497 & 3.929 & 3.346 & 2.740 & 1.897 & 0.992 &  0.077 &        \\
7.67 &       &       &       & 4.657 & 4.099 & 3.528 & 2.940 & 2.302 & 1.393 &  0.496 & $-$0.427 \\
8.00 &       &       &       &       &       &       & 3.217 & 2.632 & 1.996 &  1.103 &  0.216 \\
\enddata
\end{deluxetable*}

\begin{deluxetable*}{p{2cm}<{\centering}rrrrrrrrr}
\tablecaption{Decimal logarithm of qLFP conductive opacity in $\mathrm{cm^2\,g^{-1}}$ for He. \label{tab:He}}
\tablehead{
\hspace{1cm} & \multicolumn{7}{c}{$\log\rho (\mathrm{g\,cm^{-3}})$} \\
\multicolumn{1}{c}{$\log T (\mathrm{K})$} &   \multicolumn{1}{c}{1.0} &   \multicolumn{1}{c}{1.5} & \multicolumn{1}{c}{2.0} &   \multicolumn{1}{c}{2.5} &   \multicolumn{1}{c}{3.0} &   \multicolumn{1}{c}{3.5} &    \multicolumn{1}{c}{4.0} &    \multicolumn{1}{c}{4.5} &    \multicolumn{1}{c}{5.0}
}
\startdata
5.00 & 1.591 &       &       &       &       &       &        &        &        \\
5.50 & 2.426 & 1.558 &       &       &       &       &        &        &        \\
6.00 & 3.285 & 2.394 & 1.524 & 0.573 &       &       &        &        &        \\
6.50 & 3.883 & 3.264 & 2.369 & 1.502 & 0.583 &       &        &        &        \\
7.00 & 4.308 & 3.721 & 3.120 & 2.370 & 1.486 & 0.613 & $-$0.351 &        &        \\
7.25 & 4.507 & 3.934 & 3.348 & 2.741 & 1.927 & 1.057 &  0.156 &        &        \\
7.50 & 4.695 & 4.130 & 3.557 & 2.966 & 2.344 & 1.479 &  0.617 & $-$0.300 &        \\
7.75 &       &       & 3.754 &       & 2.584 & 1.914 &  1.041 &  0.178 & $-$0.759 \\
\enddata
\end{deluxetable*}

Compared with the general-purpose model of \cite{cassisi2007}, our qLFP calculations offer improved opacities mainly in a narrow $\rho$ -- $T$ domain corresponding to partially degenerate electrons and moderate Coulomb coupling.
Historically, the accuracy of models for the conductive opacity in this regime have been limited by how well $e-e$ scattering is accounted for.
The relative importance of $e-e$ versus $e-i$ scattering scales roughly as $Z^{-1}$ \citep{braginskii1958,simakov2014}.
$e-e$ collisions are thus most important to the conductive opacity of low-$Z$ elements, where they contribute roughly equally as $e-i$ ones.
An inaccurate model for $e-e$ scattering thus does not much affect the conductive opacity of metals, but it can severely affect that of H and He.

  The influence of $e-e$ scattering on the conductive opacity enters in two distinct ways.
  These can be understood by decomposing the conductive opacity as
  \begin{equation}
    \label{eq:decomp}
    \kappa_c = S \kappa_{c,ei} + \kappa_{c,ee}
  \end{equation}
  where $\kappa_{c,ei}$ and $\kappa_{c,ee}$ are, respectively, the conductive opacities obtained by considering only $e-i$ and $e-e$ collisions \citep{desjarlais2017}.
  The factor $S$ represents the indirect modification of the $e-i$ scattering term due the presence of $e-e$ collisions.
  This arises because the electron distribution function takes on a different shape (and thus has a different associated heat flux) depending on whether or not $e-e$ collisions occur.
  The assumption $S=1$ corresponds to Matthiessen's rule for conduction: that  $e-i$ and $e-e$ are totally independent scattering mechanisms and their respective thermal resistivities can be added ``in series'' \citep{matthiessen1858}.
  In kinetic theory, this limit is obtained only in the lowest-order Chapman--Enskog approximation to the thermal conductivity \citep{lampe1968b,hubbard1969}.
  Such an approximation is reasonable for degenerate or high-$Z$ plasmas, but it can lead to large errors in partially degenerate H or He.
  For instance, \cite{desjarlais2017} have shown that for partially degenerate and nondegenerate H, the factor $S$ takes values of about $0.6-0.7$.
  In terms of a correction to $\kappa_{c,ei}$, this reshaping effect from $S$ is of comparable importance in H as including the direct effect from $\kappa_{c,ee}$.
  Our new conductive opacities differ seriously from those of \cite{cassisi2007} due to an improved account of both the direct and indirect effects in the partially degenerate regime.

  In \citeauthor{cassisi2007}, the conductive opacity is constructed from Matthiessen's rule
\begin{equation}
  \label{eq:opac-cassisi}
  \kappa_c \approx \kappa_{c,ei} + \kappa_{c,ee}
  ,
\end{equation}
with $\kappa_{c,ei}$ and $\kappa_{c,ee}$ being treated as completely independent.
That is, the reshaping effect from $S$ in Equation~\eqref{eq:decomp} is not considered.
Since $S \le 1$, the neglect of this reshaping effect is to systematically overestimate the relative importance of $\kappa_{c,ei}$ versus $\kappa_{c,ee}$ in determining the overall opacity.
Indeed, as is shown in Section~\ref{sec:implementation}, our new qLFP conductive opacities are systematically smaller than those of \citeauthor{cassisi2007}

The direct $e-e$ contribution, $\kappa_{c,ee}$, is treated by \citeauthor{cassisi2007} with an interpolation between the degenerate and classical limits.
The interpolation reproduces the results of \cite{hubbard1969} at conditions of weak Coulomb coupling.
The regime of moderate Coulomb coupling and partial electron degeneracy is then either not constrained by theory or reproduces \citeauthor{hubbard1969}'s values on the verge of that model's breakdown.
Due to the use of mean-force scattering potentials, our qLFP calculations are more accurate at higher Coulomb coupling than \citeauthor{hubbard1969}'s model and provide the first predictive theory to accurately treat this small but important region of white dwarf phase space.

We did not extend our calculations to elements heavier than H and He as such calculations would not be applicable to white dwarf models. The qLFP theory is superior to the calculations of \cite{cassisi2007} only in a $\rho - T$ domain that corresponds to the envelopes of white dwarfs, which are made of H and/or He. One notable exception however are the Hot DQ white dwarfs, which have C-dominated atmospheres and envelopes \citep{dufour2007,dufour2008}. Still, improved conductive opacities for C-rich envelopes would be of limited applicability, as the evolution of Hot DQs below $T_{\rm eff} = 18{,}000\,$K---where much of the impact of the new conductive opacities occur (see Section~\ref{sec:evol_cool})---remains unclear. C has likely largely settled down by then, transforming Hot DQs into DQs with He-dominated atmospheres \citep{coutu2019}, but no detailed evolutionary calculations exist at the moment. Additionally, current uncertainties on the composition of the envelopes of Hot DQs (in particular, the He and O abundances) prevent any accurate modeling and, in any case, the qLFP theory should be in better agreement with the \citeauthor{cassisi2007} results for C than for H or He due to the reduced importance of $e-e$ collisions relative to $e-i$ ones with increasing $Z$.

For completeness, we note that the error function in Equation~\eqref{eq:lnL-ee} recommended by \cite{shaffer2020b} is physically motivated but not prescribed by the qLFP theory.
Its purpose is to roll off the constant $5/4$ term in the Coulomb logarithm which is important at high temperatures but incorrect at low temperatures.
Another physically plausible functional form for the roll off was considered
\begin{equation}
  \operatorname{erf}[(2T/3T_F)^3] \to (1 + e^{\mu_e/k_BT})^{-1}
\end{equation}
While this slightly changes the values of the conductive opacities at moderate degeneracy, it does not result in any substantive change to the evolution sequences presented in Section~\ref{sec:implications} which would change our conclusions.

\section{Implementation of the new conductive opacities}
\label{sec:implementation}
Most modern white dwarf evolution codes rely on the conductive opacities of \cite{cassisi2007}. Those opacity tables have the advantage of spaning a wide range of density and temperature conditions, making them applicable to the whole structures of white dwarfs. Our new opacities affect a narrow region of the whole $\rho-T$ domain covered by the \citeauthor{cassisi2007} tables. Therefore, it would be useful to have a way to keep using the \citeauthor{cassisi2007} opacities across the structure of white dwarfs, but to correct them in the moderately coupled and moderately degenerate regime where the qLFP theory is expected to be superior. To do so, we designed analytic functions to smoothly correct the \citeauthor{cassisi2007} tables where appropriate.

\begin{figure*}
  \centering
  \includegraphics[height=7cm]{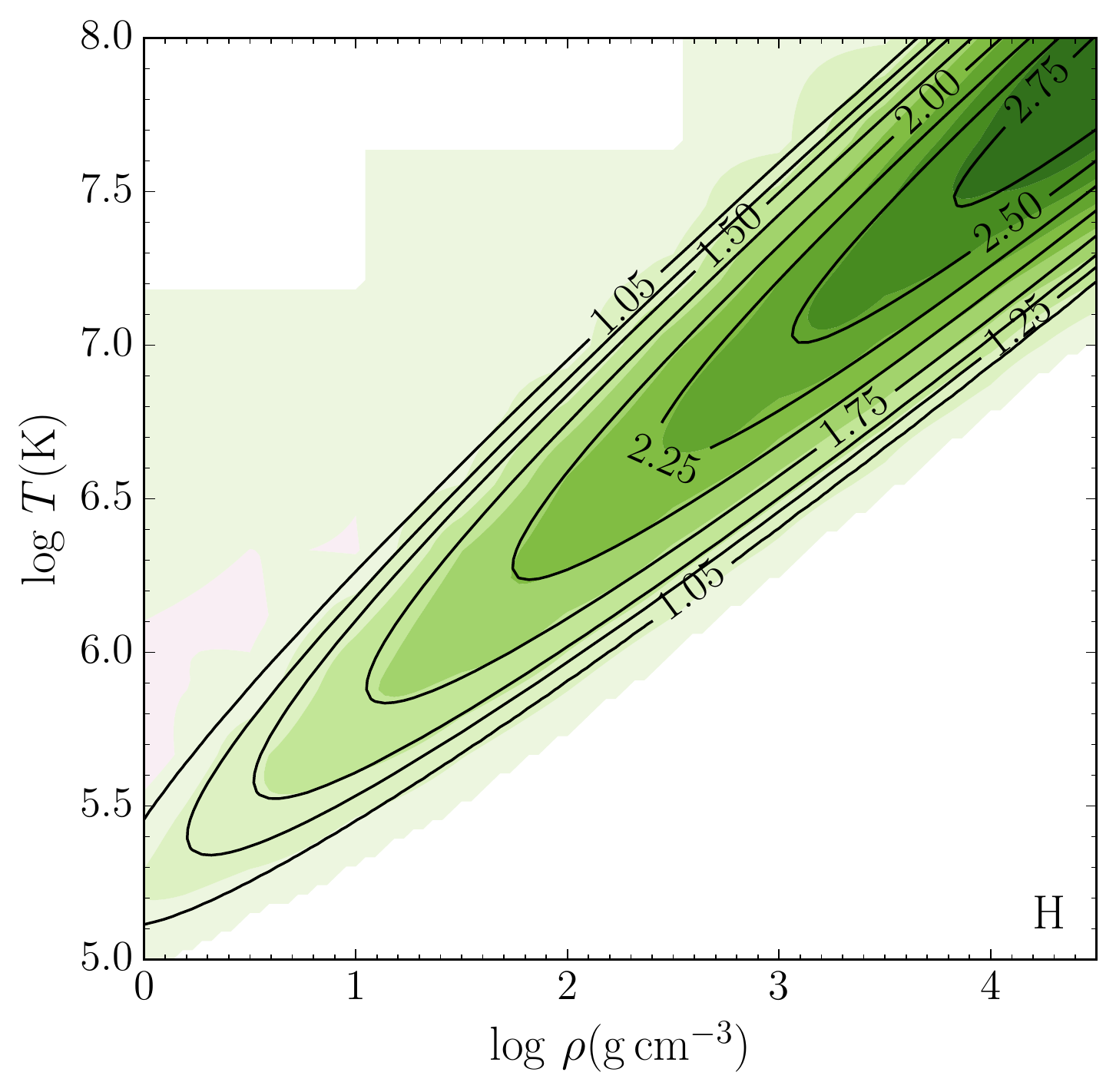}
  \includegraphics[height=7cm]{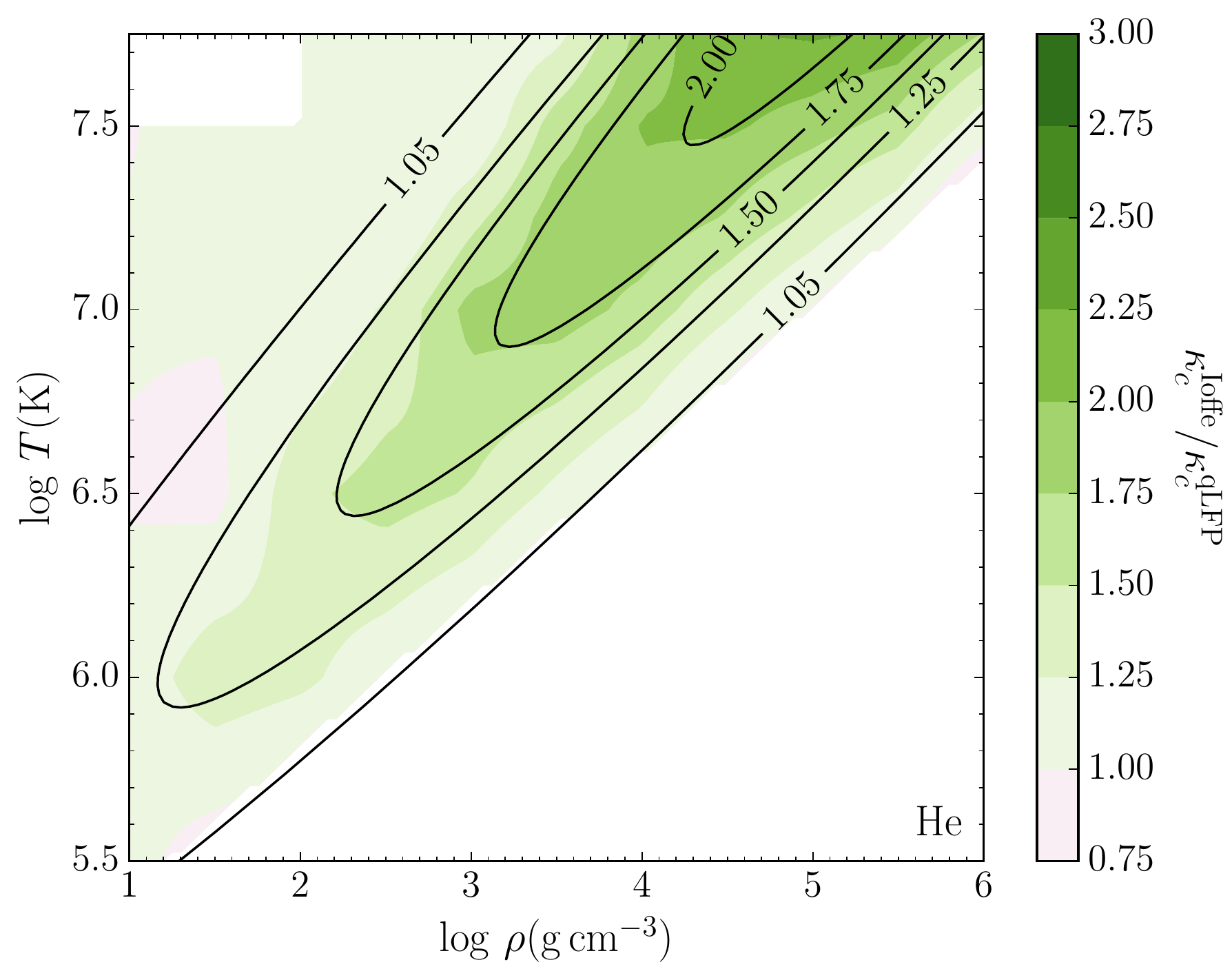}
  \caption{Comparison between our analytic corrections (contour lines) and the direct ratio of the \citeauthor{cassisi2007} and qLFP opacities (color map). Our results are shown for both H (left panel) and He (right panel). The color maps were limited to the $\rho-T$ region where the qLFP theory is expected to be superior.}
  \label{fig:fit}
\end{figure*}

We correct the \citeauthor{cassisi2007} opacities ($\kappa_c^{\rm Ioffe}$) using
\begin{equation}
\kappa_c^{\rm qLFP} (\rho, T) = \frac{\kappa_c^{\rm Ioffe} (\rho,T)}{1 + g(\rho^*, T^*)\,H \left[ g (\rho^*, T^*) \right]},
\label{eq:fitmaster}
\end{equation}
where
\begin{equation}
\begin{split}
g(\rho^*,T^*) = a \exp \left[ - \frac{\left( T^* \cos \alpha  + \rho^* \sin \alpha  \right)^2}{\sigma_T^2} \right. \\ 
\left. - \frac{ \left(T^* \sin \alpha  - \rho^*  \cos \alpha \right)^2}{\sigma_\rho^2} \right] ,
\end{split}
\end{equation}
and where $\rho^* = \log \rho/ \rho_0$, $T^* = \log T / T_0$, and
\begin{equation}
H \left[ g (\rho^*, T^*) \right] = 0.5 \tanh \{ b \left[ g(\rho^*,T^*) - 0.5 \right] \} + 0.5.
\label{eq:Hfit}
\end{equation}
The numerical parameters $\alpha$, $a$, $b$, $\rho_0$, $T_0$, $\sigma_{\rho}$ and $\sigma_{T}$ are given in Table~\ref{tab:fit} for both H and He plasmas. Figure~\ref{fig:fit} compares this analytic correction (contour lines) to the correction obtained by directly comparing the values obtained in Section~\ref{sec:opac} to those reported in \citeauthor{cassisi2007}\footnote{We use the tables given in \url{http://www.ioffe.ru/astro/conduct/}. Note that the last update of those tables was done in July 2006. Our analytic model would need to be modified if a new update is made available in the future.} (color map). Our simple analytic functions are sufficient to reproduce the decrease in opacity predicted by our new calculations.

\begin{table}
\centering
 \caption{Numerical parameters of Equations \eqref{eq:fitmaster} to \eqref{eq:Hfit}. \label{tab:fit}}
\begin{tabular}{lrr}
\hline
\hline
Parameter & \multicolumn{1}{c}{H} & \multicolumn{1}{c}{He} \\
\hline
$\alpha$ & $-$0.52 & $-$0.46 \\ 
$a$ & 2.00 & 1.25 \\
$b$ & 10.00 & 2.50 \\
$\log \rho_0$(cm$^{-3}$) & 5.45 & 6.50 \\
$\log T_0$(K) & 8.40 & 8.57 \\
$\sigma_{\rho}$ & 5.14 & 6.20 \\
$\sigma_{T}$ & 0.45 & 0.55 \\
\hline
\end{tabular}
\end{table}

Our analytic corrections go smoothly to $\kappa_c^{\rm qLFP} = \kappa_c^{\rm Ioffe}$ outside the range of application of the qLFP theory, so there is no need to apply any density or temperature cut-off when implementing those corrections in white dwarf models. That being said, the corrections are probably off beyond the high-temperature, high-density limit of Figure~\ref{fig:fit}. In this limit, the qLFP results diverge from those of \citeauthor{cassisi2007}, so the analytic model is extrapolating our results and it is unclear when (or if) we should recover $\kappa_c^{\rm qLFP} = \kappa_c^{\rm Ioffe}$. However, in practice, this is not a problem as the conductive opacities in this $\rho-T$ region are of no interest for white dwarf envelopes (for reference, see the white dwarf structures shown in Figure~\ref{fig:rhoT_models}).

\section{Implications for white dwarf cooling}
\label{sec:implications}
In this section, we investigate how our new conductive opacities for the H/He envelope affect the cooling of white dwarfs. To do so, we implemented the analytic model presented in the previous section in STELUM, the evolution code developed by the Montreal group. The constitutive physics implemented in this code is very similar to what is outlined in \cite{fontaine2001}. An up-to-date and more complete description of STELUM will be given in Bédard et al. (in preparation). 

We assume in this paper that DA white dwarfs have an envelope thickness of $q({\rm He}) \equiv M_{\rm He} / M_{\star} = 10^{-2}$ and $q({\rm H})=10^{-4}$, and that DBs have an envelope thickness of $q({\rm He})=10^{-2}$ and $q({\rm H})=10^{-10}$. These values are consistent with those derived from evolution models and empirical constraints \citep{renedo2010,bergeron2011,koester2015,rolland2018}, although we note that variations are expected depending on the white dwarf mass and its previous evolution. We also always assume a homogeneous and equimassic C/O core.

\subsection{Behavior at high temperatures}
\label{sec:evol_hot}
\begin{figure}
  \centering
  \includegraphics[width=\columnwidth]{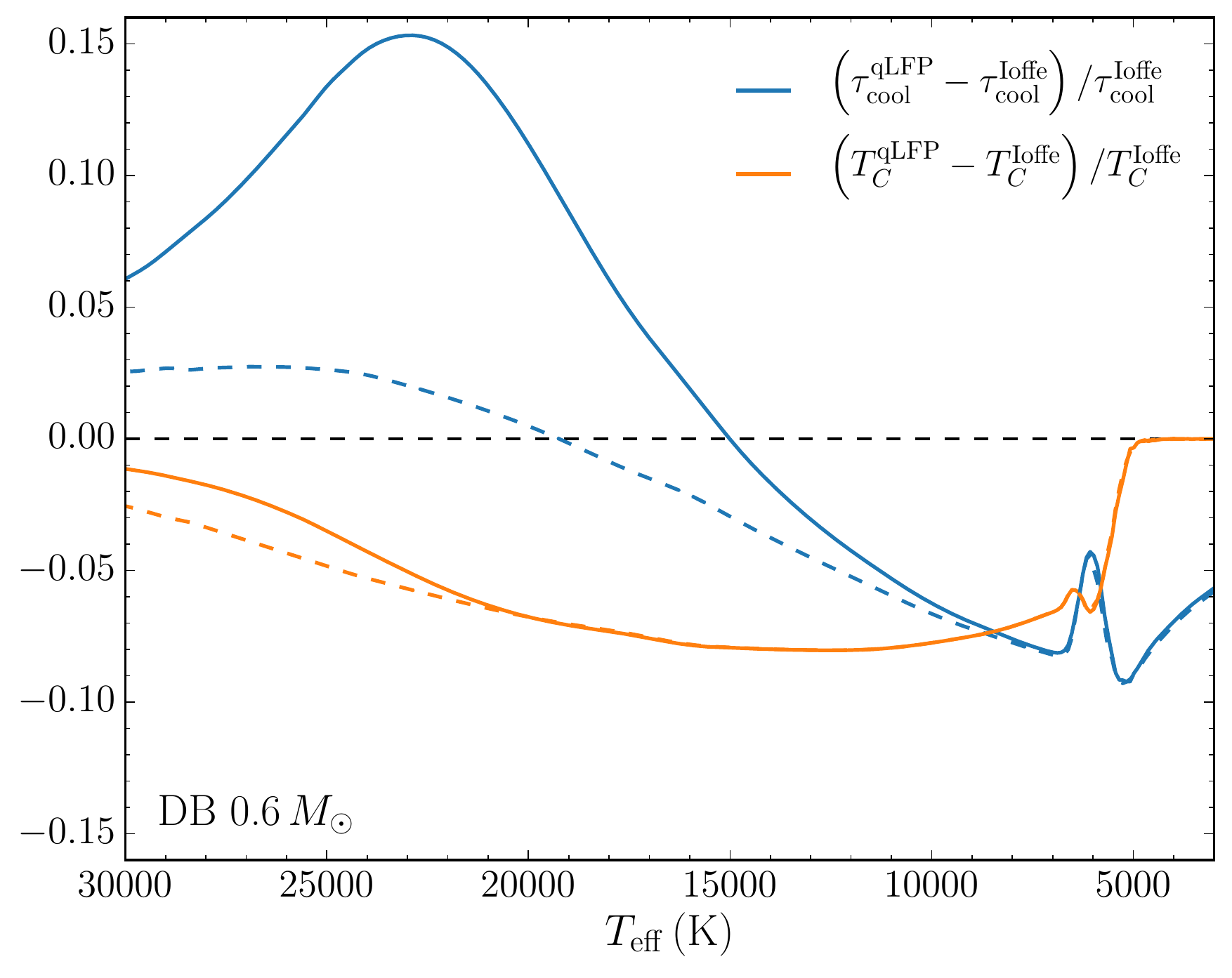}
  \caption{Relative difference of the white dwarf cooling time ($\tau_{\rm cool}$) and central temperature ($T_C$) between models that include the qLFP corrections for the conductivities in the envelope and models that directly use the \citeauthor{cassisi2007} tables. The solid lines compare sequences that include neutrino cooling and the dashed lines, sequences that omit neutrino cooling. Those results were obtained for a $0.6\,M_{\odot}$ DB white dwarf.}
  \label{fig:evol_hot}
\end{figure}

We first look at the effect of the new conductive opacities during the early phases of white dwarf cooling. Figure~\ref{fig:evol_hot} compares evolution sequences that include our corrections to the conductive opacities to sequences that directly use the \citeauthor{cassisi2007} tables for a $0.6\,M_{\odot}$ DB white dwarf. A priori, the behavior of the sequences that include the reduced opacities obtained from our new calculations is puzzling. Naively, one would assume that reduced opacities would lead to a faster cooling due to the more efficient transport of heat from the core to the surface. This is what we see at low temperatures, but the contrary occurs at $T_{\rm eff} \gtrsim 15{,}000\,{\rm K}$. 

The reason for this counterintuitive behavior is analogous to that given by \citet[see also B\'edard et al. in preparation]{salaris2013} in the context of a comparison between the opacities of \cite{itoh1983}, \cite{mitake1984} and \cite{hubbard1969} and those of \cite{cassisi2007}. A decrease in the conductive opacities implies an initial faster cooling of the core, as shown in Figure~\ref{fig:evol_hot}. With a lower core temperature, the efficiency of neutrino cooling---an important cooling process at high temperatures---is greatly reduced, which explains why the cooling time for a given effective temperature subsequently increases. To explicitly test this explanation, Figure~\ref{fig:evol_hot} also shows sequences where neutrino cooling is turned off (dashed lines). The difference in cooling times is then much smaller, which confirms our interpretation. Note that although we did not discuss it here for the sake of conciseness, a qualitatively similar behavior is obtained for DA white dwarfs.

\subsection{Behavior at low temperatures}
\label{sec:evol_cool}
While interesting and instructive, the differences in cooling times at high temperatures have a limited impact on white dwarf age dating since hot white dwarfs are very young. A 15\% difference on the cooling time at $T_{\rm eff}=18{,}000\,{\rm K}$ (Figure~\ref{fig:evol_hot}) only corresponds to a $<20\,{\rm Myr}$ age difference. In contrast, the $\approx$10\% difference seen at 5000\,K corresponds to an age difference of $>0.5\,{\rm Gyr}$, with more important implications for white dwarf cosmochronology.

\begin{figure*}
  \centering
  \includegraphics[width=2\columnwidth]{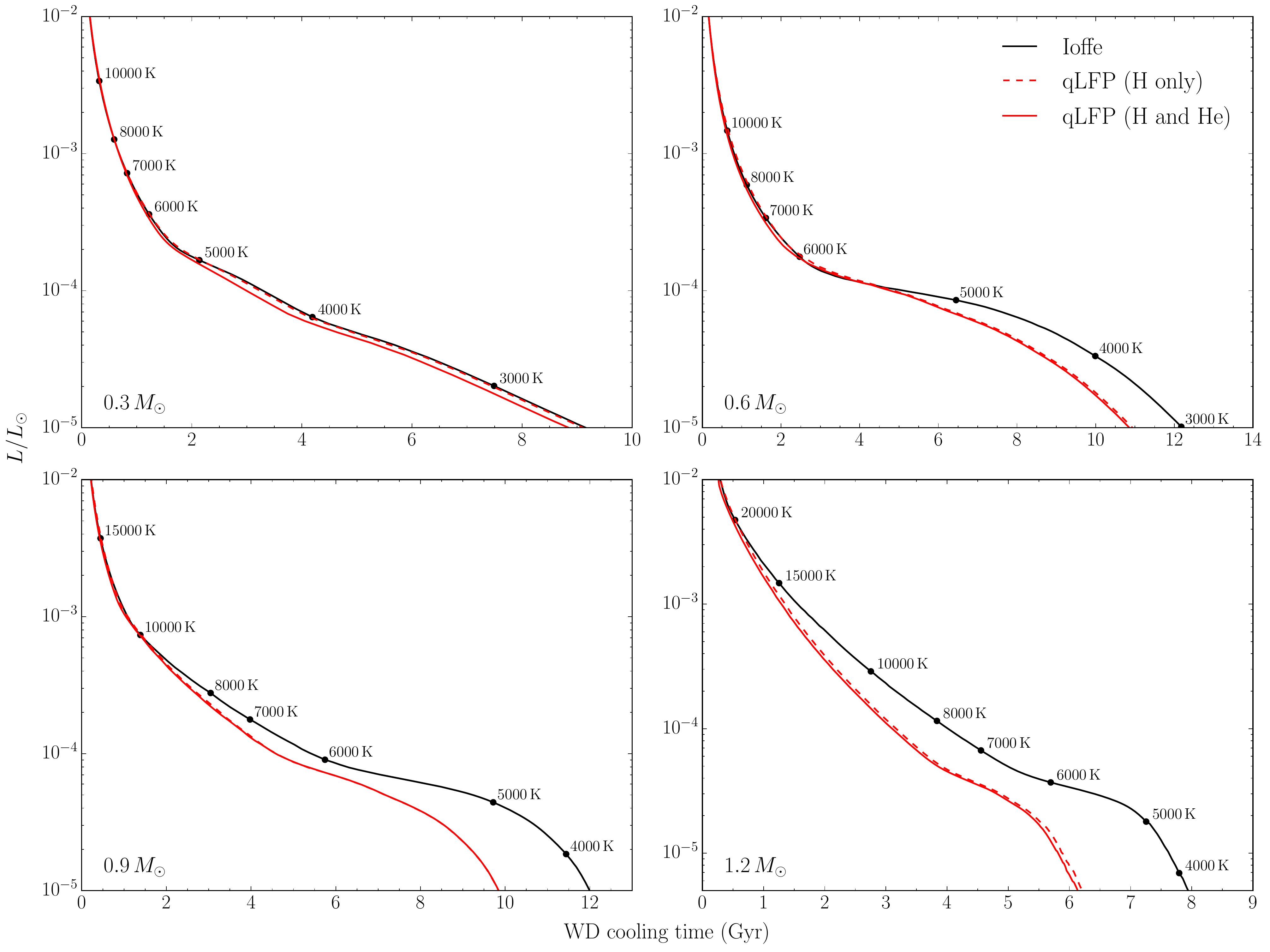}
  \caption{Cooling sequences for DA white dwarfs of different masses. Sequences that include the qLFP corrections to the envelope conductivities are shown in red (dashed red in the case where only the H conductivities are corrected) and sequences that directly use the \citeauthor{cassisi2007} values are shown in black. For reference, temperature labels along the sequences track the evolution of the effective temperature.}
  \label{fig:seq_DA}
\end{figure*}

\begin{figure*}
  \centering
  \includegraphics[width=2\columnwidth]{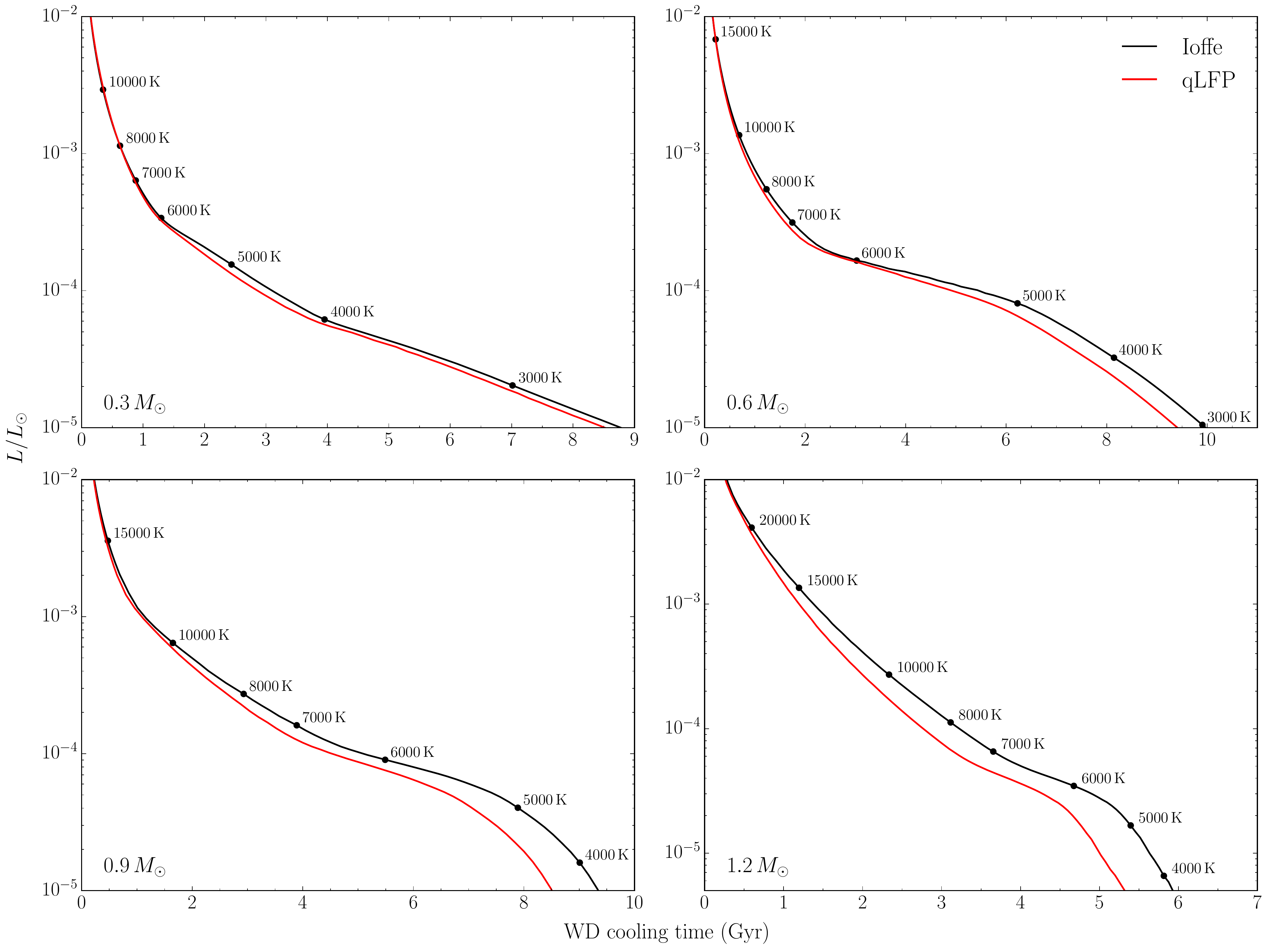}
  \caption{Same as Figure~\ref{fig:seq_DA}, but for DB white dwarfs.}
  \label{fig:seq_DB}
\end{figure*}

Figures \ref{fig:seq_DA} and \ref{fig:seq_DB} compare evolution sequences that include and omit the qLFP corrections for DA and DB white dwarfs, respectively. The new conductive opacities have a dramatic effect on white dwarf cooling, leading to age differences of up to $2\,{\rm Gyr}$ for cool white dwarfs. These important changes to the existing cooling sequences will have implications for the comparison between observational data and theoretical models. This will be discussed in Section~\ref{seq:comp_obs}. In the remainder of this section, we discuss the behavior of the cooling sequences, as shown in Figures~\ref{fig:seq_DA} and \ref{fig:seq_DB}. More specifically, we will explain why the age differences (1) grow mostly after $T_{\rm eff}\approx 6000\,{\rm K}$, (2) are larger for DA than for DB white dwarfs and (3) are larger for massive white dwarfs.

\begin{figure*}
  \centering
  \includegraphics[width=\columnwidth]{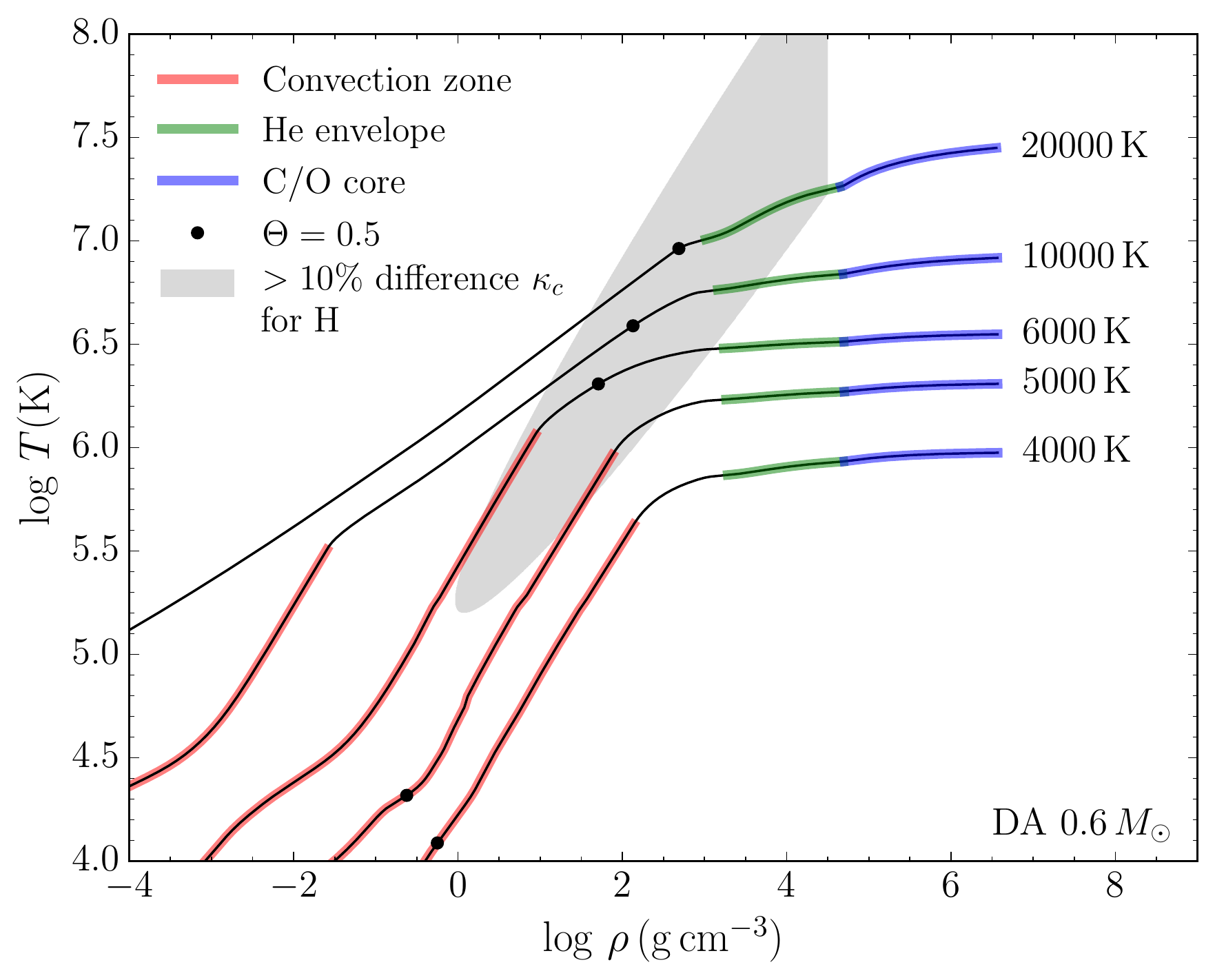}
  \includegraphics[width=\columnwidth]{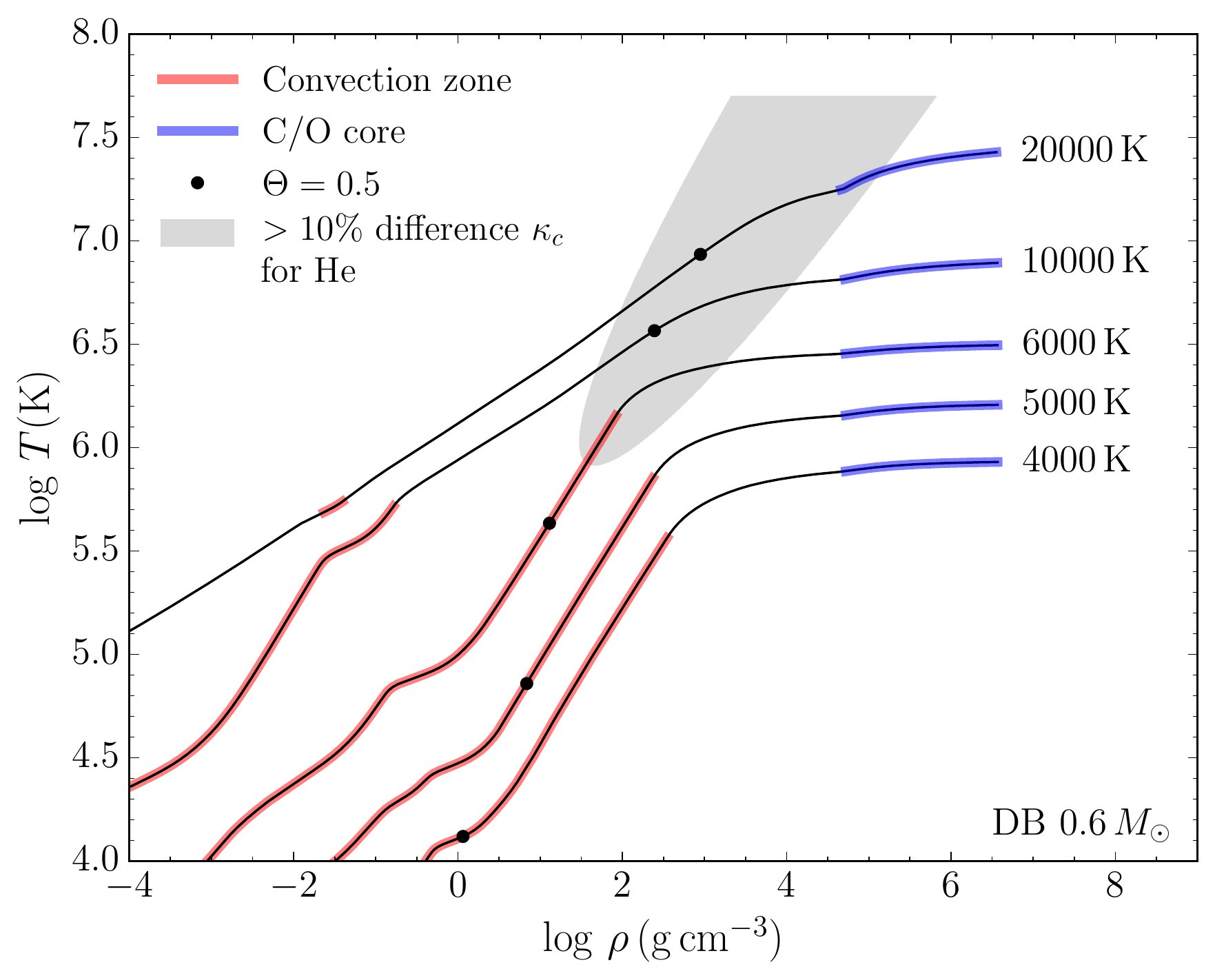}
  \caption{$\rho-T$ structures of 0.6\,$M_{\odot}$ DA (left panel) and DB (right panel) models. The black circles identify where the plasma degeneracy parameter ($\Theta = T/T_F$) becomes smaller than 0.5. Convective coupling is approximately achieved when this black circle reaches the deepest layer of the convection zone (highlighted in red). The regions shaded in gray delimit the $\rho-T$ domains where the H (left panel) and He (right panel) conductive opacities presented in this work and those of \citeauthor{cassisi2007} differ by more than 10\%.}
  \label{fig:rhoT_models}
\end{figure*}

In all cases where a significant age difference appears between sequences that include the new conductivities and those that omit them, the age differences grow mostly after $T_{\rm eff}\approx 6000\,{\rm K}$. This transition is due to convective coupling, which happens roughly around 6000\,K (the exact temperature depends on the white dwarf mass and its envelope composition; \citealt{fontaine2001}, Figure~2). Convective coupling occurs when the superficial convection zone and the degenerate and conductive layers of the envelope reach one another (Figure~\ref{fig:rhoT_models}). This leads to the rapid onset of a strong coupling between the core and the outer layers and implies that the star has to suddenly get rid of an excess of thermal energy, which temporarily slows down the cooling process. The larger the excess of thermal energy is, the more pronounced is the cooling delay due to convective coupling. Because the sequences that include the qLFP corrections cool down more efficiently prior to convective coupling (but after the phase where neutrino cooling is important, see Section~\ref{sec:evol_hot}), their core temperature is cooler by the time convective coupling is achieved. There is therefore less excess thermal energy to evacuate during the initial phase of convective coupling, leading to a smaller cooling delay. This explains the rapid age divergence observed in Figures~\ref{fig:seq_DA} and \ref{fig:seq_DB} around 6000\,{\rm K}.

A comparison of Figures~\ref{fig:seq_DA} and \ref{fig:seq_DB} reveals an important difference between DA and DB white dwarfs: the cooling sequences of DAs are more affected than those of DBs by the new conductive opacities. The main explanation for this difference can be found in Figure~\ref{fig:fit}. Since He ions are twice as charged as H ones, the conductive opacity is somewhat less sensitive to the $e-e$ scattering physics, and the qLFP opacities are closer to the \citeauthor{cassisi2007} values for He than for H, meaning that the cooling rate is more affected if a thick hydrogen envelope is present.

Both for DAs and DBs, Figures~\ref{fig:seq_DA} and \ref{fig:seq_DB} show that the more massive a star is, the more the new conductive opacities affect the cooling sequences. This is due to the fact that the envelopes of massive white dwarfs are significantly denser. Because of this, a large portion of the envelope becomes conductive earlier than for lower-mass objects, which explains the increased sensitivity to the conductive opacities. The opposite happens with low-mass objects, where the structure is shifted to lower densities. In the case of the 0.3\,$M_{\odot}$ DA white dwarf shown in Figure~\ref{fig:seq_DA}, this shift is important enough that the conductive hydrogen layers are outside the $\rho - T$ region affected by the new opacities. Only the conductivity of the He envelope is changed, which explains the observed behavior (i.e., the sequence where the qLFP results are only applied to the H layer is virtually identical to the reference sequence, while the sequence that also includes modifications to the He opacities is not).

\subsection{Implications for the comparison of theoretical cooling sequences and observational data}
\label{seq:comp_obs}
The faster cooling of white dwarfs after convective coupling described in the previous section could help improve the agreement between observational data and evolution models. Based on the analysis of the mass distribution of a large sample of DA white dwarfs, \cite{kilic2020} have recently shown that there are much fewer cool ($T_{\rm eff} \lesssim 10{,}000$\,K) and massive ($M\gtrsim1 M_{\odot}$) white dwarfs than predicted by population synthesis calculations. Massive white dwarfs must cool down and become too faint to be observable more rapidly than current cooling models predict. Our new opacities should contribute to solve this problem, as they lead to a faster cooling, especially for the more massive objects (Figure~\ref{fig:seq_DA}). However, such a comparison between population syntheses and observational data is outside the scope of this work. Moreover, as pointed out by Kilic et al., other improvements to cooling models (e.g., a more accurate treatment of phase separation during crystallization, a better understanding of the role of $^{22}$Ne diffusion) will be required before the mass distribution of DA white dwarfs can be successfully reproduced by population syntheses.

On the other hand, the faster cooling implied by our new conductivities can also be a problem. \cite{tremblay2019} have successfully reproduced the low-luminosity cut-off of the white dwarf luminosity function of massive DAs by assuming a standard 10\,Gyr age for the Galactic disk and a constant stellar formation rate. Accelerating the cooling of massive DA white dwarfs would be problematic as it would likely worsen the fit to the low-luminosity cut-off. However, we note that additional cooling delays could compensate this effect. In particular, the energy released by the sedimentation of O upon crystallization---which depends on the exact shape of the C/O phase diagram, a challenging calculation to perform (\citealt{segretain1993}, \citealt{horowitz2010}, Blouin et al. in preparation)---might have been underestimated possibly because of uncertainties on the initial C/O profile. Similarly, for some stars, diffusion of $^{22}$Ne and the associated cooling delay \citep{bildsten2001,garciaberro2008,althaus2010,camisassa2016} may be more important than currently assumed in evolution models \citep{cheng2019}.

\subsection{Impurities}
\begin{figure}
  \centering
  \includegraphics[width=\columnwidth]{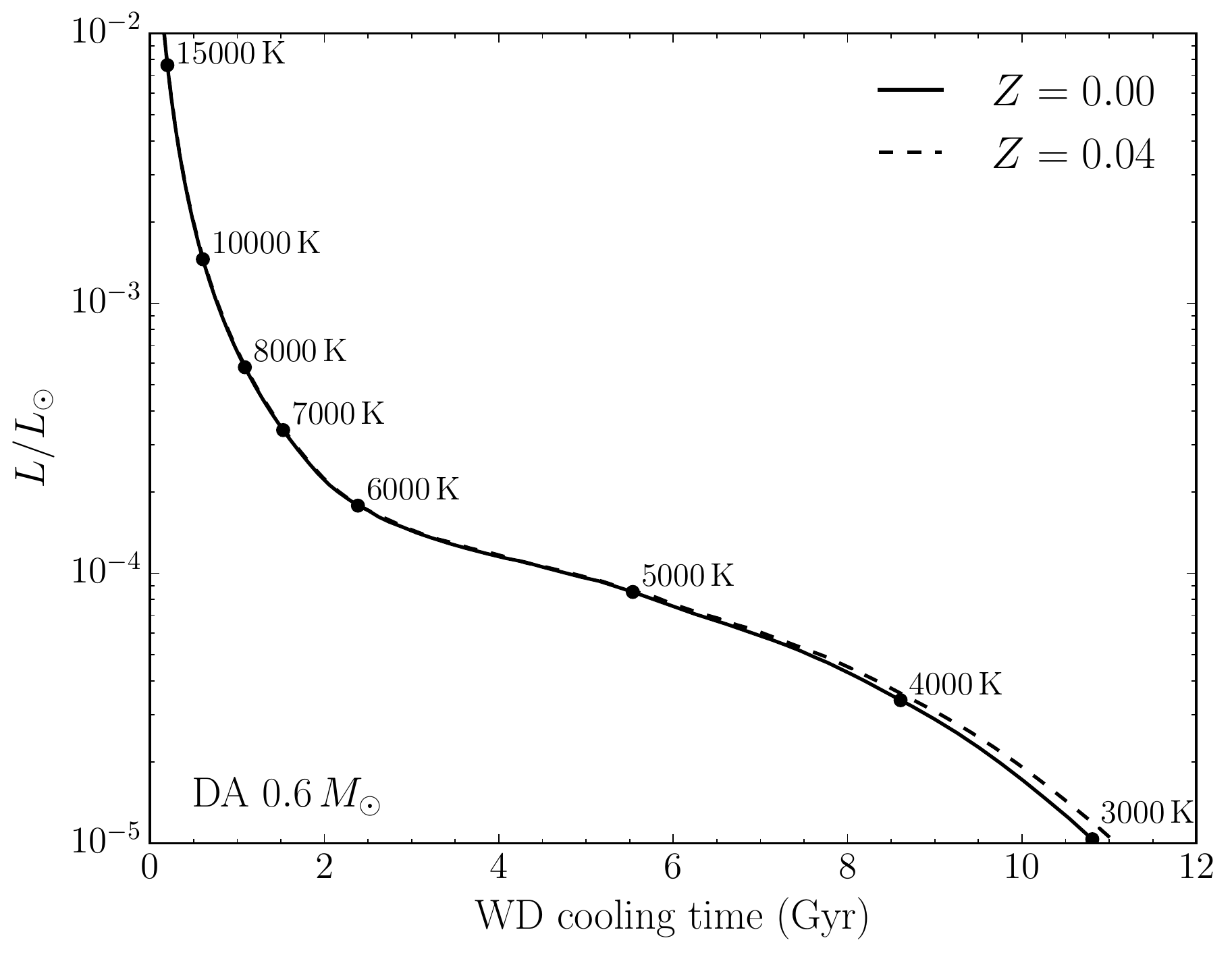}
  \caption{Comparison of cooling sequences for a 0.6\,$M_{\odot}$ DA white dwarf assuming $Z=0.00$ and $Z=0.04$ in the envelope. The metal-to-metal abundance ratios are assumed to be solar.}
  \label{fig:metallicity}
\end{figure}

One detail we have overlooked in the calculation of our conductive opacities is the likely presence of metallic impurities in the H/He envelope. To account for the effect of impurities on the conductivities, we rely on the mixing rule given in \citeauthor{cassisi2007}, where the effective charge
\begin{equation}
Z_{\rm eff} = \left( \sum_j Z_j ^2 \frac{n_j}{\sum_i n_i} \right)^{1/2}
\end{equation}
of the mixture is computed and then used to interpolate between the opacity tables of the different elements. To check if this approximation can be a problem in the context of white dwarf cooling models, we compare in Figure~\ref{fig:metallicity} two extreme models: one with a $Z=0.00$ metallicity and one with $Z=0.04$ (assuming that the metal-to-metal abundance ratios are solar). Clearly, even for those extreme metallicity values, the difference between both cooling sequences is small, which demonstrates that we do not have to worry about the exact treatment of impurities for the calculation of the conductive opacities.

\section{Conclusions}
\label{sec:conclusions}
We presented improved conductive opacities for the moderately degenerate and moderately coupled regime that characterizes an important part of the H/He envelopes of white dwarf stars. The improvement mainly comes from an accurate account of $e-e$ scattering in this regime, for which there was previously no predictive and accurate theory. The new conductive opacities are up to a factor of three smaller than those of \cite{cassisi2007}. We gave analytical fits that can be implemented in any white dwarf evolution code to correct the \citeauthor{cassisi2007} conductivities in the appropriate regime.

We have shown that the reduced conductive opacities initially lead to a slower white dwarf cooling due to the more rapid inhibition of neutrino cooling. More importantly for white dwarf cosmochronology, the new opacities lead to a much more rapid cooling at cooler temperatures (especially after convective coupling is achieved), with age differences of up to 2~Gyr at $T_{\rm eff}=4000\,{\rm K}$ for massive DA white dwarfs. This is an important effect that could help explain the recently identified depletion of massive DA white dwarfs after crystallization \citep{kilic2020}.

\acknowledgements
We thank P. Brassard for sharing with us the STELUM evolution code, which was used to test the influence of the new conductive opacities. We are also grateful to the anonymous referee for useful comments that have improved our discussion on the role played by the new conductive opacities in cooling models.

Research presented in this article was supported
by the Laboratory Directed Research and Development program of Los
Alamos National Laboratory under project number 20190624PRD2.
This work was performed under the auspices of the U.S. Department of Energy
under Contract No. 89233218CNA000001.

\bibliographystyle{aasjournal}
\bibliography{references}

\end{document}